\documentclass[12pt]{article}

\usepackage{amsmath}
\usepackage{epsfig}
\usepackage[sort&compress,square,comma]{natbib}

\begin{document}

\renewcommand{\thefootnote}{\fnsymbol{footnote}}

\begin{titlepage}

\begin{center}
 \vspace*{.8 in} {\Large\bf Designer Gene Networks: Towards
Fundamental Cellular Control}\\
 \vspace*{0.6 in} {\large Jeff Hasty$^{1}$, Farren Isaacs$^1$,
Milos Dolnik$^2$, David McMillen$^{1}$, and J.~J.~Collins$^1$}\\
 ~\\
 ~\\
 {\large August 20, 2000}\\
 \vspace*{0.6 in} {$^{1}$Center for BioDynamics and Dept.~of
Biomedical Engineering, Boston University, 44 Cummington St.,
Boston, MA 02215}\\
 \vspace*{0.2 in} {$^{2}$Dept.~of Chemistry and Center for Complex
Systems, Brandeis University, Waltham, MA 02454}\\
 \vspace*{0.4 in} Submitted to {\em Chaos}

\end{center}

\end{titlepage}

\newpage

\noindent ABSTRACT \hspace{0.1in} The engineered control of
cellular function through the design of synthetic genetic networks
is becoming plausible. Here we show how a naturally occurring
network can be used as a parts list for artificial network design,
and how model formulation leads to computational and analytical
approaches relevant to nonlinear dynamics and statistical physics.
We first review the relevant work on synthetic gene networks,
highlighting the important experimental findings with regard to
genetic switches and oscillators. We then present the derivation
of a deterministic model describing the temporal evolution of the
concentration of protein in a single-gene network. Bistability in
the steady-state protein concentration arises naturally as a
consequence of autoregulatory feedback, and we focus on the
hysteretic properties of the protein concentration as a function
of the degradation rate. We then formulate the effect of an
external noise source which interacts with the protein degradation
rate. We demonstrate the utility of such a formulation by
constructing a protein switch, whereby external noise pulses are
used to switch the protein concentration between two values.
Following the lead of earlier work, we show how the addition of a
second network component can be used to construct a relaxation
oscillator, whereby the system is driven around the hysteresis
loop. We highlight the frequency dependence on the tunable
parameter values, and discuss design plausibility. We emphasize
how the model equations can be used to develop design criteria for
robust oscillations, and illustrate this point with parameter
plots illuminating the oscillatory regions for given parameter
values. We then turn to the utilization of an intrinsic cellular
process as a means of controlling the oscillations. We consider a
network design which exhibits self-sustained oscillations, and
discuss the driving of the oscillator in the context of
synchronization. Then, as a second design, we consider a synthetic
network with parameter values near, but outside, the oscillatory
boundary. In this case, we show how resonance can lead to the
induction of oscillations and amplification of a cellular signal.
Finally, we construct a toggle switch from positive regulatory
elements, and compare the switching properties for this network
with those of a network constructed using negative regulation. Our
results demonstrate the utility of model analysis in the
construction of synthetic gene regulatory networks.
\normalsize

\vspace{0.5cm}

\baselineskip=24pt

\newpage

\section*{Lead Paragraph}
{\bf Many fundamental cellular processes are governed by genetic
programs which employ protein-DNA interactions in regulating
function. Owing to recent technological advances, it is now
possible to design synthetic gene regulatory networks. While the
idea of utilizing synthetic networks in a therapeutic setting is
still in its infancy, the stage is set for the notion of
engineered cellular control at the DNA level. Theoretically, the
biochemistry of the feedback loops associated with protein-DNA
interactions often leads to nonlinear equations, and the tools of
nonlinear analysis become invaluable. Here we utilize a naturally
occurring genetic network to elucidate the construction and design
possibilities for synthetic gene regulation. Specifically, we show
how the genetic circuitry of the bacteriophage $\lambda$ can be
used to design switching and oscillating networks, and how these
networks can be coupled to cellular processes.  This work suggests
that a genetic toolbox can be developed using modular design
concepts. Such advancements could be utilized in engineered
approaches to the modification or evaluation of cellular
processes.}

\section{Introduction}
Remarkable progress in genomic research is leading to a complete
map of the building blocks of biology. Knowledge of this map is,
in turn, fueling the study of gene regulation, where proteins
often regulate their own production or that of other proteins in a
complex web of interactions. Post-genomic research will likely
center on the dissection and analysis of these complex dynamical
interactions. While the notions of protein-DNA feedback loops and
network complexity are not
new~\cite{glass1,savageau1,kauffman,glass2,glass3,savageau2,goodwin,tyson1,
ackers,palsson,moran,reinitz}, experimental advances are inducing
a resurgence of interest in the quantitative description of gene
regulation~\cite{novak1,novak2,hammond,keller,mcadams2,arkin2,smolen,wolf,
elowitz,gardner,becskei,endy,sveiczer}. These advances are
beginning to set the stage for a {\em modular} description of the
regulatory processes underlying basic cellular
function~\cite{novak1,shapiro,mcadams1,hartwell,lauffenburger,knight}.
In light of nearly three decades of parallel progress in the study
of complex nonlinear and stochastic processes, the project of
quantitatively describing gene regulatory networks is timely.

The concept of engineering genetic networks has roots that date
back nearly half a century~\cite{monod,novick}. It is relatively
recent, however, that experimental progress has made the design
and implementation of genetic networks amenable to quantitative
analysis. There are two dominant reasons for constructing
synthetic networks. First, simple networks represent a first step
towards logical cellular control, whereby biological processes can
be manipulated or monitored at the DNA level~\cite{chen}. Such
control could have a significant impact on post-genomic
biotechnology. From the construction of simple switches or
oscillators, one can imagine the design of genetic code, or
software, capable of performing increasingly elaborate
functions~\cite{bray,knight}. A second complementary motivation
for network construction is the scientific notion of reduced
complexity; the inherently reductionist approach of decoupling a
simple network from its native and often complex biological
setting can lead to valuable information regarding evolutionary
design principles~\cite{barkai}.

Ultimately, we envision the implementation of synthetic networks
in therapeutic applications. However, such a utilization depends
on concurrent progress in efforts to uncover basic genomic and
interspecies information. For example, broad applicability will
only arise with detailed information regarding tissue-specific
promoters, proteins, and genes. Likewise, quantitative network
design is contingent on a firm understanding of cellular
differentiation and fundamental processes such as transcription,
translation, and protein metabolism. More crucially, delivery is a
major hurdle; without identifiable cell-specific recognition
molecules, there is no method for introducing a network to a
specific type of cell. Since, in many regards, therapeutic
applications are somewhat premature, we focus on the
implementation of synthetic networks in less complicated
organisms. The design of synthetic circuits and optimization of
their function in bacteria, yeast, or other plant organisms should
reveal nonlinear properties that can be employed as possible
mechanisms of cellular control.

In this paper, we develop several models describing the dynamics
of the protein concentration in small self-contained synthetic
networks, and demonstrate techniques for externally controlling
the dynamics. Although our results are general, as they originate
from networks designed with common gene regulatory elements, we
ground the discussion by considering the genetic circuitry of
bacteriophage $\lambda$. Since the range of potentially
interesting behavior is wide, we focus primarily on the
concentration of the $\lambda$ repressor protein.  We first show
how bistability in the steady-state value of the repressor protein
can arise from a single-gene network. We then show how an external
noise source affecting protein degradation can be introduced to
our model, and how the subsequent Langevin equation is analyzed by
way of transforming to an equation describing the evolution of a
probability function. We then obtain the steady-state mean
repressor concentration by solving this equation in the long-time
limit, and discuss its relationship to the magnitude of the
external perturbation. This leads to a potentially useful
application, whereby one utilizes the noise to construct a genetic
switch. We next show how the addition of a second network
component can lead to a genetic relaxation oscillator.  We study
the oscillator model in detail, highlighting the essential design
criteria. We introduce a mechanism for coupling the oscillator to
a time-varying genetic process.  In the model equations, such
coupling leads to a driven oscillator, and we study the resulting
system in the framework of synchronization. We illustrate the
utility of such driving through the construction of an amplifier
for small periodic signals. Finally, we turn to the construction
of a genetic toggle switch, and compare switching times for our
network with those of a network constructed using negative
regulation.

\section{Background}
Many processes involving cellular regulation take place at the
level of gene transcription~\cite{GenesVI}. The very nature of
cellular differentiation and role-specific interaction across cell
types implicates a not yet understood order to cellular processes.
Various modeling approaches have successfully described certain
aspects of gene regulation in specific biological
systems~\cite{reinitz,novak1,novak2,arkin2,endy,sveiczer,ackers,shea,crp_cites}.
It is only recent, however, that designed network experiments have
arisen in direct support of regulatory
models~\cite{elowitz,gardner,becskei}. In this section, we
highlight the results of these experimental studies, and set the
stage for the discussion of the network designs described in this
work.

For completeness, we first discuss the basic concepts of promoters
and regulatory feedback loops~\cite{mcclure,von_hippel}. A {\em
promoter region} (or, simply, a promoter) denotes a segment of DNA
where an RNA polymerase molecule will bind and subsequently
transcribe a gene into an mRNA molecule. Thus, one speaks of a
promoter as driving the transcription of a specific gene.
Transcription begins downstream from the promoter at a particular
sequence of DNA that is recognized by the polymerase as the start
site of transcription. A chemical sequence of DNA known as the
{\em start codon} codes for the region of the gene that is
converted into amino acids, the protein building blocks. Feedback
arises when the translated protein is capable of interacting with
the promoter that drives its own production or promoters of other
genes. Such {\em transcriptional regulation} is the typical method
utilized by cells in controlling expression~\cite{jacob,dickson},
and it can occur in a positive or negative sense. Positive
regulation, or activation, occurs when a protein increases
transcription through biochemical reactions that enhance
polymerase binding at the promoter region. Negative regulation, or
repression, involves the blocking of polymerase binding at the
promoter region.  Proteins commonly exist as multi-subunits or
multimers which perform regulatory functions throughout the cell
or serve as DNA-binding proteins. Typically, protein homodimers
(or heterodimers) regulate transcription, and this fact is
responsible for much of the nonlinearity that arises in genetic
networks~\cite{smolen}.

Recently, there have been three important experimental studies
involving the design of synthetic genetic networks. All three
employ the use of repressive promoters. In order of increasing
complexity, they consist of (i) a single autorepressive promoter
utilized to demonstrate the interplay between negative feedback
and internal noise~\cite{becskei}, (ii) two repressive promoters
used to construct a genetic toggle switch~\cite{gardner}, and
(iii) three repressive promoters employed to exhibit sustained
oscillations~\cite{elowitz}. We now briefly review the key
findings in these three studies.

In the single gene study, both a negatively controlled and an
unregulated promoter were utilized to study the effect of
regulation on variations in cellular protein
concentration~\cite{becskei}. The central result is that negative
feedback decreases the cell-to-cell fluctuations in protein
concentration measurements. Although the theoretical notion of
network-induced decreased variability is not new~\cite{savageau4},
this study empirically demonstrates the phenomenon through the
measurement of protein fluorescence distributions over a
population of cells. The findings show that, for a repressive
network, the fluorescence distribution is significantly tightened,
and that such tightening is proportional to the degree to which
the promoter is negatively controlled. These results suggest that
negative feedback is utilized in cellular design as a means for
mitigating variations in cellular protein concentrations. Since
the number of proteins per cell is typically small, internal noise
is thought to be an important issue, and this study speaks to
issues regarding the reliability of cellular processes in the
presence of internal noise.

The toggle switch involves a network where each of two proteins
negatively regulates the synthesis of the other; protein ``{\em
A}'' turns off the promoter for gene ``{\em B}'', and protein {\em
B} turns off the promoter for gene {\em A}~\cite{gardner}. In this
work, it is shown how certain biochemical parameters lead to two
stable steady states, with either a high concentration of {\em A}
(low {\em B}), or a high concentration of {\em B} (low {\em A}).
Reliable switching between states is induced through the transient
introduction of either a chemical or thermal stimulus, and shown
to be significantly sharper than for that of a network designed
without co-repression. Additionally, the change in fluorescence
distributions during the switching process suggests interesting
statistical properties regarding internal noise. These results
demonstrate that synthetic toggle switches can be designed and
utilized in a cellular environment. Co-repressive switches have
long been proposed as a common regulatory theme~\cite{monod2}, and
the synthetic toggle serves as a model system in which to study
such networks.

In the oscillator study, three repressible promoters were used to
construct a network capable of producing temporal oscillations in
the concentrations of cellular proteins~\cite{elowitz}. The
regulatory network was designed with cyclic repressibility;
protein {\em A} turns off the promoter for gene {\em B}, protein
{\em B} turns off the promoter for gene {\em C}, and protein {\em
C} turns off the promoter for gene {\em A}. For certain
biochemical parameters, the ``repressilator'' was shown to exhibit
self-sustained oscillations over the entire growth phase of the
host {\em E. coli} cells. Interestingly, the period of the
oscillations was shown to be longer than the bacterial septation
period, suggesting that cellular conditions important to the
oscillator network were reliably transmitted to the progeny cells.
However, significant variations in oscillatory phases and
amplitudes were observed between daughter cells, and internal
noise was proposed as a plausible decorrelation mechanism. These
variations suggest that, in order to circumvent the effects of
noise, naturally-occurring oscillators might need some additional
form of control. Indeed, an important aspect of this study was its
focus on the utilization of synthetic networks as tools for
biological inference. In this regard, the repressilator work
provides potentially valuable information pertaining to the design
principles of other oscillatory systems, such as circadian clocks.

These studies represent important advances in the
engineering-based methodology of synthetic network design. In all
three, the experimental behavior is consistent with predictions
which arise from continuum dynamical modeling. Further,
theoretical models were utilized to determine design criteria,
lending support to the notion of an engineering-based approach to
genetic network design. These criteria included the use of strong
constitutive promoters, effective transcriptional repression,
cooperative protein interactions, and similar protein degradation
rates. In the immediate future, the construction and analysis of a
circuit containing an activating control element (i.e., a positive
feedback system) appears to be a next logical step.

In this work, we present several models describing the design of
synthetic networks in prokaryotic organisms. Specifically, we
will utilize genetic components from the virus bacteriophage
$\lambda$. While other quantitative studies have concentrated on
the switching properties of the $\lambda$ phage
circuitry~\cite{reinitz,arkin2,ackers,shea}, we focus on its value
as a parts list for designing synthetic networks. Importantly, the
biochemical reactions that constitute the control of $\lambda$
phage are very well characterized; the fundamental biochemical
reactions are understood, and the equilibrium association
constants are
known~\cite{ackers,ptashne,meyer,johnson1,johnson2,ohlendorf}. In
its naturally occurring state, $\lambda$ phage infects the
bacteria {\em Escherichia coli} ({\em E. coli}). Upon infection,
the evolution of $\lambda$ phage proceeds down one of two
pathways. The {\em lysis} pathway entails the viral destruction
of the host, creating hundreds of phage progeny in the process.
These progeny can then infect other bacteria. The {\em
lysogenous} pathway involves the incorporation of the phage DNA
into the host genome. In this state, the virus is able to
dormantly pass on its DNA through the bacterial progeny. The
extensive interest in $\lambda$ phage lies in its ability to
perform a remarkable trick; if an {\em E. coli} cell infected
with a lysogen is endangered (i.e. exposure to UV radiation), the
lysogen will quickly switch to the lysis pathway and abandon the
challenged host cell.

The biochemistry of the viral ``abandon-ship'' response is a
textbook example~\cite{GenesVI} of cellular regulation via a
naturally-occurring genetic switch. The lytic and lysogenic states
are controlled by the \textit{cro} and \textit{cI} genes,
respectively.  These genes are regulated by what are known as the
$P_{RM}$ (\textit{cI} gene) and $P_R$ (\textit{cro} gene)
promoters. They overlap in an operator region consisting of the
three binding sites OR1, OR2, and OR3, and the Cro and $\lambda$
repressor~(``repressor'', the {\em cI} product) protein actively
compete for these binding sites. When the Cro protein (product of
\textit{cro} gene) binds to these sites, it induces lysis. When
repressor binds, lysogeny is maintained and lysis suppressed. When
potentially fatal DNA damage is sensed by an {\em E. coli} host,
part of the cellular response is to attempt DNA repair through the
activation of a protein called RecA. $\lambda$ phage has evolved
to utilize RecA as a signal; RecA degrades the viral repressor
protein and Cro subsequently assumes control of the promoter
region. Once Cro is in control, lysis ensues and the switch is
thrown.

\section{Bistability in a Single-Gene Network}
In this section, we develop a quantitative model describing the
regulation of the $P_{RM}$ operator region of $\lambda$ phage. We
envision that our system is a DNA plasmid consisting of the
promoter region and {\em cI} gene.

As noted above, the promoter region contains the three operator
sites known as OR1, OR2, and OR3. The basic dynamical properties
of this network, along with a categorization of the biochemical
reactions, are as follows. The gene {\it cI} expresses repressor
(CI), which in turn dimerizes and binds to the DNA as a
transcription factor. This binding can take place at one of the
three binding sites OR1, OR2, or OR3. The binding affinities are
such that, typically, binding proceeds sequentially; the dimer
first binds to the OR1 site, then OR2, and lastly
OR3~\cite{footnote_lambda}. Positive feedback arises due to the
fact that downstream transcription is enhanced by binding at OR2,
while binding at OR3 represses transcription, effectively turning
off production and thereby constituting a negative feedback loop.

The chemical reactions describing the network are naturally
divided into two categories -- fast and slow. The fast reactions
have rate constants of order seconds, and are therefore assumed to
be in equilibrium with respect to the slow reactions, which are
described by rates of order minutes. If we let $X$, $X_2$, and $D$
denote the repressor, repressor dimer, and DNA promoter site,
respectively, then we may write the equilibrium reactions
\begin{eqnarray}
X+X & \stackrel{K_{1}}{\rightleftharpoons } & X_{2}
 \label{fast} \\
 D+X_{2} & \stackrel{K_{2}}{\rightleftharpoons } & D_1\nonumber \\
 D_1+X_{2} & \stackrel{K_{3}}{\rightleftharpoons } & D_2D_1\nonumber \\
 D_2D_1+X_{2} & \stackrel{K_{4}}{\rightleftharpoons } & D_3D_2D_1\nonumber
\end{eqnarray}
where $D_i$ denotes dimer binding to the OR{\em i} site, and the
$K_i = k_i/k_{-i}$ are equilibrium constants. We let
$K_3=\sigma_1K_2$ and $K_4=\sigma_2K_2$, so that $\sigma_1$ and
$\sigma_2$ represent binding strengths relative to the dimer-OR1
strength.

The slow irreversible reactions are transcription and degradation.
If no repressor is bound to the operator region, or if a single
repressor dimer is bound to OR1, transcription proceeds at a
normal unenhanced rate. If, however, a repressor dimer is bound to
OR2, the binding affinity of RNA polymerase to the promoter region
is enhanced, leading to an amplification of transcription.
Degradation is essentially due to cell growth. We write the
reactions governing these processes as
\begin{eqnarray}
 D + P & \overset{k_{t}}{\rightarrow} & D + P + nX
\label{slow} \\
 D_1 + P & \overset{k_{t}}{\rightarrow} & D_1 + P + nX \nonumber
\\
 D_2D_1 + P & \overset{\alpha k_{t}}{\rightarrow} & D_2D_1 + P + nX \nonumber
\\
 X & \overset{k_x}{\rightarrow} & ~ \nonumber
\end{eqnarray}
where $P$ denotes the concentration of RNA polymerase, $n$ is the
number of repressor proteins per mRNA transcript, and $\alpha>1$
is the degree to which transcription is enhanced by dimer
occupation of OR2.

Defining concentrations as our dynamical variables, $x=[X]$,
$x_2=[X_2]$, $d_0=[D]$, $d_1=[D_1]$, $d_2=[D_2D_1]$, and
$d_3=[D_3D_2D_1]$, we can write a rate equation describing the
evolution of the concentration of repressor,
\begin{equation}
 \dot x =  -2k_{1}x^2 + 2k_{-1}x_2 + nk_{t}p_{0}(d_0+d_1+\alpha d_2) - k_{x}x
  \label{xeqn}
\end{equation}
where we assume that the concentration of RNA polymerase $p_0$
remains constant during time.

We next eliminate $x_2$ and the $d_i$ from Eq.~(\ref{xeqn}) as
follows. We utilize the fact that the reactions in
Eq.~(\ref{fast}) are fast compared to expression and degradation,
and write algebraic expressions
\begin{eqnarray}
 x_{2} & = & K_{1}x^{2} \\
 d_{1} & = & K_{2}d_{0}x_{2}=(K_{1}K_{2})d_{0}x^{2}\nonumber \\
 d_{2} & = & K_{3}d_{1}x_{2}=\sigma_{1}(K_{1}K_{2})^2d_{0}x^{4}\nonumber \\
 d_{3} & = & K_{4}d_{2}x_{2}=\sigma _{1}\sigma _{2}(K_{1}K_{2})^{3}d_{0}x^{6}\nonumber
 \label{Eqn:AlgebraicRels}
\end{eqnarray}

Further, the total concentration of DNA promoter sites $d_{T}$ is
constant, so that
\begin{equation}
 md_T=d_0(1+K_1K_2x^2+\sigma_1(K_1K_2)^2x^4+\sigma_1\sigma_2(K_1K_2)^3x^6)
\end{equation}
where $m$ is the copy number for the plasmid, i.e., the number of
plasmids per cell.

We next eliminate two of the parameters by rescaling the repressor
concentration $x$ and time. To this end, we define the
dimensionless variables $\widetilde{x}=x \sqrt{K_1K_2}$ and
$\widetilde{t}=t (k_t p_0 d_T n \sqrt{K_1K_2})$. Upon substitution
into Eq.~(\ref{xeqn}), we obtain
\begin{equation}
\dot x = \frac {m(1+x^2 + \alpha\sigma_1~x^4)}
{1+x^2+\sigma_1x^4+\sigma_1\sigma_2x^6} - \gamma_x~x
\label{nodimx}
\end{equation}
where $\gamma_x = k_x/(d_Tnk_tp_0\sqrt{K_1K_2})$, the time
derivative is with respect to $\widetilde{t}$, and we have
suppressed the overbar on $x$. The equilibrium constants are
$K_1=5.0 \times 10^7$M$^{-1}$ and $K_2=3.3 \times
10^8$M$^{-1}$~\cite{ackers,ptashne,johnson1,johnson2}, so that the
transformation from the dimensionless variable x to the total
concentration of repressor~(monomeric and dimeric forms) is given
by [CI]$=(7.7x+3.0x^2)$~nM. The scaling of time involves the
parameter $k_t$, and since transcription and translation are
actually a complex sequence of reactions, it is difficult to give
this lump parameter a numerical value. However, in
Ref.~\cite{hasty_mcmillen}, it is shown that, by utilizing a model
for the lysogenous state of the $\lambda$ phage, a consistency
argument yields a value for the product of parameters
$(d_tnk_tp_0)=87.6$~nM~min$^{-1}$. This leads to a transformation
from the dimensionless time $\widetilde{t}$ to time measured in
minutes of $t(minutes)=0.089\widetilde{t}$.

Since equations similar to Eq.~(\ref{xeqn}) often arise in the
modeling of genetic circuits~(see Refs.~\cite{focus_refs}) of this
Focus Issue), it is worth noting the specifics of its functional
form. The first term on the right hand side of Eq.~(\ref{nodimx})
represents production of repressor due to transcription. The even
polynomials in $x$ occur due to dimerization and subsequent
binding to the promoter region. As noted above, the $\sigma_i$
prefactors denote the relative affinities for dimer binding to OR1
versus that of binding to OR2~($\sigma_1$) and OR3~($\sigma_2$).
The prefactor $\alpha>1$ on the $x^4$ term is present because
transcription is enhanced when the two operator sites OR1 and OR2
are occupied~($x^2x^2$). The $x^6$ term represents the occupation
of all three operator sites, and arises in the denominator because
dimer occupation of OR3 inhibits polymerase binding and shuts off
transcription.

For the operator region of $\lambda$ phage, we have $\sigma_1 \sim
2$, $\sigma_2 \sim 0.08$, and $\alpha \sim
11$~\cite{ackers,ptashne,johnson1,johnson2}, so that the
parameters $\gamma_x$ and $m$ in Eq.~(\ref{nodimx}) determine the
steady-state concentration of repressor. The parameter $\gamma_x$
is directly proportional to the protein degradation rate, and in
the construction of artificial networks, it can be utilized as a
tunable parameter. The integer parameter $m$ represents the number
of plasmids per cell. While this parameter is not accessible
during an experiment, it is possible to design a plasmid with a
given copy number, with typical values in the range of 1-100.

The nonlinearity of Eq.~(\ref{nodimx}) leads to a bistable regime
in the steady state concentration of repressor, and in Figure 1A
we plot the steady-state concentration of repressor as a function
of the parameter $\gamma_x$. The bistability arises as a
consequence of the competition between the production of $x$ along
with dimerization and its degradation. For certain parameter
values, the initial concentration is irrelevant, but for those
that more closely balance production and loss, the final
concentration is determined by the initial value.

Before turning to the next section, we make one additional
observation regarding the synonymous issues of the general
applicability of a synthetic network and experimental measurement.
In experimental situations, a Green Fluorescent Protein~(GFP) is
often employed as a measurement tag known as a reporter gene. This
is done by inserting the gene encoding GFP adjacent to the gene of
interest, so that the reporter protein is produced in tandem with
the protein of interest. In the context of the formulation given
above, we can generalize Eq.~(\ref{nodimx}) to include the
dynamics of the reporter protein,
\begin{eqnarray}
 \dot x &=& f(x) - \gamma_x~x
 \label{nodimxg} \\
 \dot g &=& f(x) - \gamma_g~g \nonumber
\end{eqnarray}
where $f(x)$ is the nonlinear term in Eq.~(\ref{nodimx}),
$\gamma_g = k_g/(d_Tnk_tp_0\sqrt{K_1K_2})$, and the GFP
concentration is scaled by the same factor as
repressor~($\widetilde{g}=g \sqrt{K_1K_2}$). In analogy with the
equation for $x$, $k_g$ is the degradation rate for GFP, and we
have assumed that the number of proteins per transcript $n$ is the
same for both processes. This ability to co-transcribe two genes
from the same promoter and transcribe in tandem has two important
consequences. First, since proteins are typically very stable, it
is often desirable to substantially increase their degradation
rate in order to access some nonlinear
regime~\cite{elowitz,gardner}. Such a high degradation rate
typically will lead to a low protein concentration, and this, in
turn, can induce detection problems. The utilization of a GFP-type
reporter protein can help to mitigate this problem, since its
degradation rate can be left at a relatively low value. Second,
and perhaps more importantly, are the significant implications for
the generality of designer networks; in prokaryotic organisms,
{\em any} protein can be substituted for GFP and co-transcribed,
so that one network design can be utilized in a myriad of
situations.



\section{A Noise-Based Protein Switch}
We now focus on parameter values leading to bistability, and
consider how an external noise source can be utilized to alter the
production of protein. Physically, we take the dynamical variables
$x$ and $g$ described above to represent the protein
concentrations within a colony of cells, and consider the noise to
act on many copies of this colony. In the absence of noise, each
colony will evolve identically to one of the two fixed points, as
discussed above. The presence of a noise source will at times
modify this simple behavior, whereby colony-to-colony fluctuations
can induce novel behavior.

Noise in the form of random fluctuations arises in biochemical
networks in one of two ways. As discussed elsewhere in this Focus
Issue~\cite{arkin_chaos}, {\em internal} noise is inherent in
biochemical reactions, often arising due to the relatively small
numbers of reactant molecules. On the other hand, {\em external}
noise originates in the random variation of one or more of the
externally-set control parameters, such as the rate constants
associated with a given set of reactions. If the noise source is
small, its effect can often be incorporated {\em post hoc} into
the rate equations. In the case of internal noise, this is done in
an attempt to recapture the lost information embodied in the
rate-equation approximation. But in the case of external noise,
one often wishes to introduce some new phenomenon where the
details of the effect are not precisely known.  In either case,
the governing rate equations are augmented with additive or
multiplicative stochastic terms. These terms, viewed as a random
perturbation to the deterministic picture, can induce various
effects, most notably, switching between potential attractors
(i.e., fixed points, limit cycles, chaotic
attractors)~\cite{horsthemke}.

In previous work, the effects of coupling between an external
noise source and both the basal production rate and the
transcriptional enhancement process were examined~\cite{hasty}.
Here, we analyze the effect of a noise source which alters protein
degradation. Since the mathematical formulation is similar to
that of Ref.~\cite{hasty}, our goal here is to reproduce the
phenomenology of that work under different assumptions. As in
Ref.\cite{hasty}, we posit that the external noise effect will be
small and can be treated as a random perturbation to our existing
treatment; we envision that events induced will be interactions
between the external noise source and the protein degradation
rate, and that this will translate to a rapidly varying protein
degradation embodied in the external parameters $\gamma_x$ and
$\gamma_g$. In order to introduce this effect, we generalize the
model of the previous section such that random fluctuations enter
Eq.~(\ref{nodimxg}) multiplicatively,
\begin{eqnarray}
 \dot{x} &=& f(x) - (\gamma_x-\xi_x(t))x \label{langevinx} \\
 \dot{g} &=& f(x) - (\gamma_g-\xi_g(t))g \label{langeving}
\end{eqnarray}
where the $\xi_i(t)$ are rapidly fluctuating random terms with
zero mean ($<\xi_i(t)>=0$). In order to encapsulate the
independent random fluctuations, we make the standard requirement
that the autocorrelation be ``$\delta$-correlated'', i.e., the
statistics of the $\xi_i(t)$ are such that
$<\xi_i(t)\xi_j(t')>=D\delta_{i,j}(t-t')$, with $D$ proportional
to the strength of the perturbation, and we have assumed that the
size of the induced fluctuations is the same for both proteins.

Since, in Eqs.~(\ref{langevinx}) and (\ref{langeving}), the
reporter protein concentration $g$ does not couple to the equation
for the repressor concentration, the qualitative behavior of the
set of equations may be obtained by analyzing $x$. We first define
a change of variables which transforms the multiplicative Langevin
equation to an additive one. Letting $x=e^z$,
Eq.~(\ref{langevinx}) becomes,
\begin{eqnarray}
\dot z &=& \frac{1+e^{2z}+22e^{4z}}{e^z+e^{3z}+2e^{5z}+.16e^{7z}}
 -\gamma_x + \xi_x(t) \label{langevinz}\\
~    &\equiv& g(z) + \xi_x(t) \nonumber
\end{eqnarray}

Eq.~(\ref{langevinz}) can be rewritten as:
\begin{equation}
\dot{z} = -\frac{\partial \phi(z)}{\partial z} + \xi_x(t)
\label{landscape}
\end{equation}
where the potential $\phi(z)$ is introduced:
\begin{equation}
\phi(z) = -\int g(z)dz
\end{equation}
$\phi(z)$ can be viewed as an ``energy landscape'', whereby $z$ is
considered the position of a particle moving in the landscape. One
such landscape is plotted in Fig.~1B. Note that the stable fixed
points correspond to the minima of the potential $\phi$ in
Fig.~1b, and the effect of the additive noise term is to cause
random kicks to the particle (system state point) lying in one of
these minima. On occasion, a sequence of kicks may enable the
particle to escape a local minimum and reside in a new valley.

In order to analyze Eq.~(\ref{landscape}), one typically
introduces the probability distribution $P(z,t)$, which is
effectively the probability of finding the system in a state $z$
at time $t$. Then, given Eq.~(\ref{landscape}), a
Fokker-Planck~(FP) equation for $P(z,t)$ can be
constructed~\cite{vankampen}.  The steady-state solution for this
equation is given by
\begin{equation}
P_s(z)=A e^{- \frac{2}{D} \phi(z)}
\end{equation}
where $A$ is a normalization constant determined by requiring the
integral of $P_s(z)$ over all $z$ be unity.

Using the steady-state distribution, the steady-state mean~(ssm),
$<z>_{ss}$, is given by
\begin{equation}
<z>_{ss} = \int_{0}^{\infty} z A e^{- \frac{2}{D} \phi(z)} dz
\label{uss}
\end{equation}
In Fig.~1C, we plot the ssm value of $z$ as a function of $D$,
obtained by numerically integrating Eq.~(\ref{uss}).  It can be
seen that the ssm of z increases with $D$, corresponding to the
increasing likelihood of populating the upper state in Fig.~1B.

Figure~1C indicates that the external noise can be used to control
the ssm concentration. As a candidate application, consider the
following protein switch. Given parameter values leading to the
landscape of Fig.~1B, we begin the switch in the ``off'' position
by tuning the noise strength to a very low value. This will cause
a high population in the lower state, and a correspondingly low
value of the concentration. Then at some time later, consider
pulsing the system by increasing the noise to some large value for
a short period of time, followed by a decrease back to the
original low value. The pulse will cause the upper state to become
populated, corresponding to a concentration increase and a
flipping of the switch to the ``on'' position. As the pulse
quickly subsides, the upper state remains populated as the noise
is not of sufficient strength to drive the system across either
barrier (on relevant time scales). To return the switch to the off
position, the upper-state population needs to be decreased to a
low value. This can be achieved by applying a second noise pulse
of intermediate strength. This intermediate value is chosen large
enough so as to enhance transitions to the lower state, but small
enough as to remain prohibitive to upper-state transitions.

Figure~1D depicts the time evolution of the switching process for
noise pulses of strengths $D=1.0$ and $D=0.1$. Initially, the
concentration begins at a level of $\sim 0.4~\mu$M, corresponding
to a low noise value of $D=0.01$.  At $40$ minutes, a noise burst
of strength $D=1.0$ is used to drive the concentration to a value
of $\sim 2.2~\mu$M. Following this burst, the noise is returned to
its original value. At $80$ minutes, a second noise burst of
strength $D=0.1$ is used to return the concentration to its
original value.

\section{A Genetic Relaxation Oscillator}
The repressillator represents an impressive step towards the
generation of controllable {\em in vivo} genetic oscillations.
However, there were significant cell-to-cell variations,
apparently arising from small molecule number
fluctuations~\cite{elowitz,barkai}. In order to circumvent such
variability, the utilization of hysteresis-based oscillations has
recently been proposed~\cite{barkai}. In this work, it was shown
how a model circadian network can oscillate reliably in the
presence of internal noise. In this section, we describe an
implementation of such an oscillator, based on the repressor
network of Section 3.

The hysteretic effect in Fig.~1A can be employed to induce
oscillations, provided we can couple the network to a slow
subsystem that effectively drives the parameter $\gamma_x$.  This
can be done by inserting a repressor protease under the control of
a separate $P_{RM}$ promoter region.  The network is depicted in
Fig.~2A. On one plasmid, we have the network of Section 3; the
repressor protein CI, which is under the control of the promoter
$P_{RM}$, stimulates its own production at low concentrations and
shuts off the promoter at high concentrations. On a second
plasmid, we again utilize the $P_{RM}$ promoter region, but here
we insert the gene encoding the protein RcsA. The crucial
interaction is between RcsA and CI; RcsA is a protease for
repressor, effectively inactivating its ability to control the
$P_{RM}$ promoter region~\cite{rozanov}.

The equations governing this network can be deduced from
Eq.(\ref{nodimx}) by noting the following. First, both RcsA and
repressor are under the control of the same promoter, so that the
functional form of the production term $f(x)$ in Eq.(\ref{nodimx})
will be the same for both proteins. Second, we envision our
network as being constructed from two plasmids -- one for
repressor and one for RcsA, and that we have control over the
number of plasmids per cell~(copy number) of each type.  Lastly,
the interaction of the RcsA and repressor proteins leads to the
degradation of repressor. Putting these facts together, and
letting $y$ denote the concentration of RcsA, we have
\begin{eqnarray}
 \dot x &=& m_xf(x)- \gamma_x x - \gamma_{xy}xy \label{nodimxy} \\
  ~&=& m_xf(x) - \gamma(y)x \nonumber \\
 \dot y &=& m_yf(x)- \gamma_y y \nonumber
\end{eqnarray}
where $\gamma(y)\equiv \gamma_x + \gamma_{xy}y$, and $m_x$ and
$m_y$ denote the plasmid copy numbers for the two species.

In Fig.~2B, we present simulation results for the concentration of
repressor as a function time.  The nature of the oscillations can
be understood using Fig.~1A. Suppose we begin with a parameter
value of $\gamma(y)=4$ and on the upper branch of the figure.  The
large value of repressor will then serve to activate the promoter
for the RcsA, and thus lead to its increased production.  An
increase in the RcsA acts as an additive degradation term for
repressor (see Eq.~\ref{nodimxy}), and thus effectively induces
slow motion to the right on the upper branch of Fig.~1A.  This
motion will continue until the repressor concentration falls off
the upper branch at $\gamma(y) \sim 5.8$.  At this point, with the
repressor concentration at a very low value, the promoters are
essentially turned off.  Then, as RcsA begins to degrade, the
repressor concentration slowly moves to left along the lower
branch of Fig.~1A, until it encounters the bifurcation point at
$\gamma(y) \sim 3.6$. It then jumps to its original high value,
with the entire process repeating and producing the oscillations
in Fig.~2B.

The oscillations in Fig.~2B are for specific parameter values; of
course, not all choices of parameters will lead to oscillations.
The clarification of the specific parameter values leading to
oscillations is therefore important in the design of synthetic
networks~\cite{elowitz}. For proteins in their native state, the
degradation rates $\gamma_x$ and $\gamma_y$ are very small,
corresponding to the high degree of stability for most proteins.
For example, a consistency argument applied to a similar model for
$\lambda$ phage switching~\cite{hasty_mcmillen} leads to $\gamma_x
\sim 0.004$. However, using a temperature-sensitive variety of the
repressor protein, $\gamma_x$ can be made tunable over many orders
of magnitude. Other techniques, such as SSRA tagging or titration,
can be employed to increase the degradation rate for RcsA.  The
copy numbers $m_x$ and $m_y$ can be chosen for a particular
design, and the parameter $\gamma_{xy}$, which measures the rate
of repressor degradation by RcsA, is unknown.

In Fig.~3A, we present oscillatory regimes for Eq.~(\ref{nodimxy})
as a function of $\gamma_x$ and $\gamma_y$, and for two fixed
values of the parameter $\gamma_{xy}$. We see that the oscillatory
regime is larger for smaller values of the parameter
$\gamma_{xy}$. However, the larger regime corresponds to larger
values of the degradation rate for RcsA. Interestingly, if we take
the native (i.e., without tuning) degradation rates to be
$\gamma_x \sim \gamma_y \sim 0.005$, we note that the system is
naturally poised very near the oscillatory regime.  In Fig.~3B, we
present the oscillatory regime as a function of $\gamma_x$ and
$\gamma_{xy}$, and for two fixed values of $\gamma_y$. The regime
is increased for smaller values of $\gamma_y$, and, in both cases,
small values of $\gamma_x$ are preferable. Moreover, both Figs.~3A
and 3B indicate that the system will oscillate for {\em
arbitrarily} small values of the repressor degradation parameter
$\gamma_x$. In Fig.~3C we depict the oscillatory regime as a
function of the copy numbers $m_x$ and $m_y$, and for fixed
degradation rates. Importantly, one can adjust the periodic regime
to account for the unknown parameter $\gamma_{xy}$. Figure 3C
indicates that, for oscillations, one should choose as large a
copy number as possible for the plasmid containing the repressor
protein~($m_x$). Correspondingly, one should design the RcsA
plasmid with a significantly smaller copy number $m_y$.

We now turn briefly to the period of the oscillations. If designed
genetic oscillations are to be utilized, an important issue is the
dependence of the oscillation period on the parameter values. In
Fig.~4A we plot the oscillation period for our CI-RcsA network as
a function of the degradation parameter $\gamma_y$, and for other
parameter values corresponding to the lower wedge of Fig.~3A. We
observe that an increase in $\gamma_y$ will decrease the period of
oscillations. Further, since the cell-division period for {\em E.
coli} is $\sim 35-40$ minutes, we note that the lower limit
roughly corresponds to this period, and that, at the upper limit,
we can expect four oscillations per cell division. The utilization
of tuning the period of the oscillations to the cell-division time
will be discussed in the next section. In Fig.~4B, we plot the
period as a function of the copy number $m_x$. We observe that the
period depends very weakly on the copy number.

\section{Driving the Oscillator}
We next turn to the utilization of an intrinsic cellular process
as a means of controlling the oscillations described in the
previous section. We will first consider a network design which
exhibits self-sustained oscillations~(i.e., with parameters that
are in one of the oscillatory regions of Fig.~3), and discuss the
driving of the oscillator in the context of synchronization. As a
second design, we will consider a synthetic network with parameter
values near, but outside, the oscillatory boundary.  In that case,
we will show how resonance can lead to the induction of
oscillations and amplification of a cellular signal.

We suppose that an intrinsic cellular process involves
oscillations in the production of protein $U$, and that the
concentration of $U$ is given by $u=u_0 \sin \omega t$. In order
to couple the oscillations of $U$ to our network, we imagine
inserting the gene encoding repressor adjacent to the gene
encoding $U$. Then, since $U$ is being transcribed periodically,
the co-transcription of repressor will lead to an oscillating
source
    term in Eq.~(\ref{nodimxy}),
\begin{eqnarray}
 \dot x &=& m_xf(x)- \gamma_x x - \gamma_{xy}xy + \Gamma\sin(\omega t) \label{nodimxyosc} \\
 \dot y &=& m_yf(x)- \gamma_y y \nonumber
\end{eqnarray}

We first consider parameter values as in Fig.~2, so that the
concentrations $x$ and $y$ oscillate in the absence of driving.
Here, we are interested in how the drive affects the ``internal''
oscillations. Although there are many interesting properties
associated with driven nonlinear equations such as
Eq.~(\ref{nodimxyosc}), we focus on the conditions whereby the
periodic drive can cause the dynamics to shift the internal
frequency and entrain to the external drive frequency $\omega$. We
utilize the numerical bifurcation and continuation package CONT
\cite{cont} to determine the boundaries of the major resonance
regions. These boundaries are depicted in the parameter-space plot
of Fig.~5A, where the period of the drive is plotted versus the
drive amplitude. The resonance regions form the so-called Arnold
tongues, which display an increasing range of the locking period
as the amplitude of drive is increased. Without the periodic
drive, the period of the autonomous oscillations is equal to 14.6
minutes. As one might expect, the dominant Arnold tongue is found
around this autonomous period. Within this resonance region, the
period of the oscillations is entrained, and is equal to the
external periodic force. The second largest region of frequency
locking occurs for periods of forcing which are close to half of
the period of the autonomous oscillations. As a result of the
periodic driving, we observe 1:2 locking, whereby the system
responds with one oscillatory cycle, while the drive has undergone
two cycles. Other depicted resonant regions (3:2, 2:1, 5:2, 3:1)
display significantly narrower ranges for locking periods. This
suggests that higher order frequency locking will be less common
and probably unstable in the presence of noise. Outside the
resonance regions shown in Fig.~5A one can find a rich structure
of very narrow M:N locking regions with M and N quite large,
together with quasiperiodic oscillations. The order of resonances
along the drive period axis is given by the Farey sequence \cite
{farey}, i.e. in between two resonance regions characterized by
rational numbers, M$_1$:N$_1$ and M$_2$:N$_2$, there is a region
with ratio (M$_1$+M$_2$):(N$_1$+N$_2$).





The preceding notions correspond to the driving of genetic
networks which are intrinsically oscillating. We now turn to a
network designed with parameter values just outside the
oscillatory region, and consider the use of resonance in the
following application. Suppose there is a cellular process that
depends critically on oscillations of a given amplitude. We seek a
strategy for modifying the amplitude of this process if, for some
reason, it is too small. For concreteness, consider a cellular
process linked to the cell-division period of the host for our
synthetic network. For {\em E. coli} cells at a temperature of
$\sim 37$ degrees C, this period is of order $35-40$ minutes.
Using Figs.~3A and 4A, we can deduce parameter values that will
cause a CI-RscA network to oscillate, when driven, with this
period. The lower wedge of Fig.~3A implies that, for
$\gamma_{xy}=0.1$, we should design the network with values of
$\gamma_x$ and $\gamma_y$ just below the lower boundary of the
wedge. Fig.~4A implies that, for $\gamma_x=0.1$, a choice of
$\gamma_y=0.004$ will yield oscillations with a period close to
the cell-division period. In order to stay outside the oscillatory
region, we therefore choose $\gamma_y$ just below this value.
Taken together, these choices will yield a network whereby
oscillations can be induced by cellular processes related to cell
division. In Fig.~6A, we plot the drive versus response
amplitudes~($\Gamma$ vs $\Gamma_x$) obtained from numerical
integration of Eq.~(\ref{nodimxyosc}). We see that, depending on
the proximity to the oscillatory region, oscillations are
triggered when the drive reaches some critical amplitude. In
Fig.~6B, we plot the gain $g \equiv (\Gamma+\Gamma_x)/\Gamma$ as a
function of the drive amplitude $\Gamma$, and observe that, for
certain values of the amplitude of the drive, the network can
induce a significant gain.

\section{Harnessing the Lambda Switch}
The ability to switch between multiple stable states is a critical
first step towards sophisticated cellular control schemes.
Nonlinearities giving rise to two stable states suggest the
possibility of using these states as digital signals to be
processed in cellular-level computations (see, for example,
\cite{bray,knight}). One may eventually be able to produce systems
in which sequences of such switching events are combined to
control gene expression in complex ways. In any such application,
the speed with which systems make transitions between their stable
states will act as a limiting factor on the time scales at which
cellular events may be controlled. In this section, we describe a
bistable switch based on the mechanism used by \( \lambda  \)
phage, and show that such a system offers rapid switching times.

The genetic network of \( \lambda  \) phage switches its host
bacterium from the dormant lysogenous state to the lytic growth
state in roughly twenty minutes~\cite{ptashnebook}). As discussed
in Section~2, the regulatory network implementing this
exceptionally fast switch has two main features: two proteins~(CI
and Cro) compete directly for access to promoter sites; and one of
the proteins (CI) positively regulates its own level of
transcription. Here, we compare a synthetic switch based on the \(
\lambda  \) phage's switching mechanism to another two-protein
switch~(the toggle switch described in Ref. \cite{gardner}), and
numerically show that the \( \lambda \)-like system offers a
faster switching time under comparable conditions.

To implement the synthetic \( \lambda  \) switch, we use the
plasmid described in Section~3, on which the \( P_{RM} \) promoter
controls the expression of the \( \lambda  \) repressor protein,
CI. To this, we add a second plasmid on which the \( P_{R} \)
promoter is used to control the expression of Cro. The operator
regions OR1, OR2, and OR3 exist on each plasmid, and both proteins
are capable of binding to these regions on either of the plasmids.
On the \( P_{RM} \)-promoter plasmid, transcription of CI takes
place whenever there is no protein~(of either type) bound to OR3;
when CI is bound to OR2, the rate of CI transcription is enhanced.
On the \( P_{R} \)-promoter plasmid, Cro is transcribed only when
operator site OR3 is either clear, or has a Cro dimer bound to it;
either protein being bound to either OR1 or OR2 has the effect of
halting the transcription of Cro.

Letting \( y \) represent the concentration of Cro, the
competition for operator sites leads to equations of the form \(
\dot{x}=f(x,y)-\gamma _{x}x \), \( \dot{y}=g(x,y)-\gamma _{y}y \).
We derive the form of these equations by following the process
described in Section~3. As with the CI plasmid of that section, we
have Eq.~1 describing the equilibrium reactions for the binding of
CI to the various operator sites. To these, we add the reactions
entailing the binding of Cro, and the reactions in which both
proteins are bound simultaneously to different operator sites:

\begin{eqnarray}
Y+Y & \stackrel{K_{3}}{\rightleftharpoons } & Y_{2}
\label{math-nonumber} \\ D+Y_{2} &
\stackrel{K_{4}}{\rightleftharpoons } & D^{Y}_{3}\nonumber \\
D^{Y}_{3}+Y_{2} & \stackrel{\beta _{1}K_{4}}{\rightleftharpoons }
& D^{Y}_{3}D^{Y}_{2}\nonumber \\ D^{Y}_{3}+Y_{2} & \stackrel{\beta
_{2}K_{4}}{\rightleftharpoons } & D^{Y}_{3}D^{Y}_{1}\nonumber \\
D^{Y}_{3}D^{Y}_{2}+Y_{2} & \stackrel{\beta
_{3}K_{4}}{\rightleftharpoons } &
D^{Y}_{3}D^{Y}_{2}D^{Y}_{1}\nonumber \\ D^{X}_{2}D^{X}_{1}+Y_{2} &
\stackrel{\beta _{4}K_{4}}{\rightleftharpoons } &
D^{Y}_{3}D^{X}_{2}D^{X}_{1}\nonumber \\ D^{X}_{1}+Y_{2} &
\stackrel{\beta _{5}K_{4}}{\rightleftharpoons } &
D^{Y}_{3}D^{X}_{1},\nonumber \label{Eqn:Switching:Equil}
\end{eqnarray}
where \( Y \) represents the Cro monomer, and \( D^{p}_{i} \)
represents binding of protein \( p \) to the OR{\em i} site. For
the operator region of $\lambda$ phage, we have \( \beta
_{1}\simeq \beta _{2}\simeq \beta _{3}\sim 0.08 \), and \( \beta
_{4}\simeq \beta _{5}\sim 1 \) ~\cite{ptashne,johnson1,johnson2}

The transcriptional processes are as follows. Transcription of
repressor takes place when there is no protein (of either type)
bound to OR3. When repressor is bound to OR2, the rate of
repressor transcription is enhanced, and Cro is transcribed only
when OR3 is either vacant, or has a Cro dimer bound to it. If
either repressor or Cro is bound to either OR1 or OR2, the
production of Cro is halted. These processes, along with
degradation, yield the following irreversible reactions,

\begin{eqnarray}
D+P & \stackrel{k_{tx}}{\rightarrow } & D+P+n_{x}X \\
 D^{X}_{1}+P & \stackrel{k_{tx}}{\rightarrow } & D^{X}_{1}+P+n_{x}X\nonumber \\
D^{X}_{2}D^{X}_{1}+P & \stackrel{\alpha k_{tx}}{\rightarrow } &
D^{X}_{2}D^{X}_{1}+P+n_{x}X\nonumber \\ D+P &
\stackrel{k_{ty}}{\rightarrow } & D+P+n_{y}Y\nonumber \\
D^{Y}_{3}+P & \stackrel{k_{ty}}{\rightarrow } &
D^{Y}_{3}+P+n_{y}Y\nonumber \\ X & \stackrel{k_{dx}}{\rightarrow }
&~\nonumber \\ Y & \stackrel{k_{dy}}{\rightarrow } & ~ \nonumber
\label{Eqn:Switching:Irrev}
\end{eqnarray}

Following the rate equation formulation of section 3, we obtain
\begin{eqnarray}
 \dot{x}&=&\frac{m_{x}(1+x^{2}+\alpha \sigma_{1}x^{4})}{Q(x,y)}-\gamma
 _{x}x \label{Eqn:x_dot} \\
 \dot{y}&=&\frac{m_{y}\rho_{y}(1+y^{2})}{Q(x,y)}-\gamma _{y}y \nonumber
\end{eqnarray}
\noindent where
\[
Q(x,y)=1+x^{2}+\sigma _{1}x^{4}+\sigma _{1}\sigma
_{2}x^{6}+y^{2}+(\beta _{1}+\beta _{2})y^{4}+\beta _{1}\beta
_{3}y^{6}+\sigma _{1}\beta _{4}x^{4}y^{2}+\beta _{5}x^{2}y^{2}.\]

\noindent The derivatives are with respect to dimensionless time,
with scaling as in Section~3; \(
\tilde{t}=t(k_{tx}p_{0}d_{T}n_{x}\sqrt{K_{1}K_{2}}) \), where \(
k_{tx} \) is the transcription rate constant for CI, and \( n_{x}
\) is the number of CI monomers per mRNA transcript. The integers
\( m_{x} \) and \( m_{y} \) represent the plasmid copy numbers for
the two species; \( \rho _{y} \) is a constant related to the
scaling of \( y \) relative to \( x \). The parameters \( \gamma
_{x} \) and \( \gamma _{y} \) are directly proportional to the
decay rates of CI and Cro, respectively; we will tune these values
to cause transitions between stable states. The system exhibits
bistability over a wide range of parameter values, and we plot the
null-clines in Fig.~7A.

For comparison, we now consider the co-repressive toggle switch
briefly reviewed in section~2~\cite{gardner}. This switch uses the
CI and Lac proteins, where each protein shuts off transcription
from the other protein's promoter region. The experimental design
was guided by the model equations,
\begin{eqnarray}
 \dot{u}&=&\frac{\alpha _{1}}{1+v^{\delta }}-u  \label{Eqn:u_dot} \\
 \dot{v}&=&\frac{\alpha _{2}}{1+\left[ \frac{u}{\left(
1+\frac{[IPTG]}{K}\right) ^{\eta }}\right] ^{\mu }}-kv \nonumber
\end{eqnarray}
\noindent where \( u \) and \( v \) are dimensionless
concentrations of the Lac and CI proteins, respectively, and the
time derivatives are with respect to a dimensionless time: \( \tau
=k_{d}t \), with \( k_{d}=2.52 h^{-1} \)~\cite{reinitz,arkin2}
being the protein decay rate. The dimensionless parameters \(
\alpha _{1} \), \( \alpha _{2} \), \( \delta  \), and \( \mu  \)
define the basic model. The CI protein used in the experiments is
temperature-sensitive, changing its rate of degradation with
temperature\cite{villaverde,lowman}; we modify the original model
slightly to include the factor \( k \), which represents a
varying decay rate for the CI protein. Switching is induced
either by changing \( k \), or by adjusting the concentration of
isopropyl-\( \beta \)-D-thiogalactopyranoside (IPTG); the
parameters \( K=2.9618\times 10^{-5} \) M and \( \eta =2.0015 \)
define the effect of the inducer molecule IPTG on the Lac
protein. Over a wide range of parameter values, the system has
two stable fixed points; the null-clines are shown in Figure~7B.

The time courses of switching between stable states in the two
models are shown in Figure~8; transitions are induced by
eliminating the bistability, then restoring it. The precise time
course of switching from one stable state to another is determined
by the way in which the model parameters are adjusted to eliminate
the bistability. In each case, some parameter is increased until
the system passes through a saddle node bifurcation: two stable
fixed points and one unstable fixed point collapse into a single
stable point. In an effort to examine the behaviour of the two
systems under analogous conditions, we eliminate the bistability
in every case by setting the system to a parameter value 10\%
past the bifurcation point.

The transitions shown in Figure~8 are generated as follows. The
system begins~(0-1 hour) in its default bistable state, sitting
at one of the two stable fixed points. Then~(1--4 hours) the
bistability is eliminated (as described above), with the only
remaining fixed point being such that the other protein has a high
concentration. Once the concentrations have switched, the default
parameters are restored and the system moves to the nearby stable
fixed point~(4--7 hours). Finally, the system is rendered
monostable again~(7--10 hours), causing another transition,
followed by a period~(10--11 hours) during which the bistable
parameters are restored.

Under the conditions shown, the \( \lambda  \) switch model
displays significantly more rapid transitions between its stable
states than those seen in the model of the toggle switch. The
numerical results indicate that the properties of the \( \lambda
\) switch do offer an advantage in terms of the speed of
transitions, indicating that it may be fruitful to study
synthetic models based on this natural system. Future analytical
work on models such as the one presented in this section may allow
us to make more precise statements regarding the source of this
advantage.

\section{Conclusion}
From an engineering perspective, the control of cellular function
through the design and manipulation of gene regulatory networks is
an intriguing possibility. Current examples of potential
applicability range from the use of genetically engineered
microorganisms for environmental cleanup
purposes~\cite{szafranski}, to the flipping of genetic switches in
mammalian neuronal cells~\cite{harding}. While the experimental
techniques employed in studies of this nature are certainly
impressive, it is clear that reliable theoretical tools would be
of enormous value. On a strictly practical level, such techniques
could potentially reduce the degree of ``trial-and-error''
experimentation.  More importantly, computational and theoretical
approaches will lead to testable predictions regarding the current
understanding of complex biological networks.

While other studies have centered on certain aspects of
naturally-occurring genetic regulatory
networks~\cite{reinitz,novak1,novak2,arkin2,endy,sveiczer,ackers,shea,crp_cites},
an alternative approach is to focus on the design of synthetic
networks. Such an engineering-based approach has significant
technological implications, and will lead, in a complementary
fashion, to an enhanced understanding of biological design
principles. In this work, we have shown how several synthetic
networks can be designed from the genetic machinery of the virus
$\lambda$ phage. We have highlighted some of the possible behavior
of these networks through the discussion of the design of two
types of switches and a relaxation oscillator. Additionally, in
the case of the oscillator, we have coupled the network to an
existing cellular process. Such coupling could lead to possible
strategies for entraining or inducing network oscillations in
cellular protein levels, and prove useful in the design of
networks that interact with cellular processes that require
precise timing.

With regard to model formulation, there are several intriguing
areas for further work. For one, the number of molecules governing
the biochemistry of genetic networks is often relatively small,
leading to interesting issues involving internal noise. Recent
pivotal work~\cite{mcadams2,arkin2,mcadams1} has led to a
systematic modeling approach which utilizes a Monte Carlo-type
simulation of the biochemical reactions~\cite{gillespie}. While
this approach is impressively complete, its complexity makes
analysis nearly impossible. An alternative approach could entail
the use of Langevin equations, whereby the effects of internal
noise are incorporated into stochastic terms whose magnitudes are
concentration-dependent. Indeed, in the context of genetic
switches, this approach has recently been
suggested~\cite{bialek}. The advantage of this formulation is
that stochastic effects can be viewed as a perturbation to the
deterministic picture, so that analytic tools can be utilized.


A potentially important technical issue involves the implicit
assumption that the reactions take place in three-dimensional
space. While this assumption is perhaps the most natural, proteins
have been observed sliding along a DNA molecule in search of a
promoter region~\cite{harada}, so that protein-DNA reactions might
effectively take place on a surface. While this would not alter
the qualitative form of Eq.~(\ref{nodimx}), the exponents on the
variable $x$ could take on other values~\cite{savageau3}, and
this, in turn, could lead to significant quantitative differences.

It has been nearly 30 years since the pioneering theoretical work
on interacting genetic networks~\cite{glass1}-\cite{tyson1}. Due,
in part, to the inherent complexity of regulatory networks, the
true significance of these studies had to await technological
advances. Current progress in the study of both naturally
occurring and synthetic genetic networks suggests that, as the
pioneers envisioned, tools from nonlinear dynamics and statistical
physics will play important roles in the description and
manipulation of the dynamics underlying cellular control.

~~\\ ~~\\

\noindent ACKNOWLEDGEMENTS. We warmly acknowledge insightful
discussions with William Blake, Michael Elowitz, Doug Mar, John
Reinitz, and John Tyson. This work is supported by The Fetzer
Institute~(J.~H.) and the Office of Naval Research.

\newpage
\clearpage

\section*{Figure Captions}
\noindent FIG.~1. Results for additive noise with parameter value
$m=1$. (A) Bifurcation plot for the steady-state concentration of
repressor versus the model parameter $\gamma_x$. (B) The energy
landscape. Stable equilibrium values of Eq.~(\ref{langevinz})
(with $D=0$) correspond to the valleys at $z=-1.6$ and $0.5$, with
an unstable value at $z=-0.52$. (B) Steady-state probability
distributions for noise strengths of $D=0.04$~(solid line) and
$D=0.4$~(dotted line). (C) The steady-state equilibrium value of
$z$ plotted versus noise strength. The corresponding concentration
will increas as the noise causes the upper state of (B) to become
increasingly populated. (D) Simulation of Eqs.~(\ref{langevinx})
and (\ref{langeving}) demonstrating the utilization of external
noise for protein switching. Initially, the concentration begins
at a level of $\mbox{[GFP]}\sim 0.4~\mu$M corresponding to a low
noise value of $D=0.01$. After $40$ minutes, a large $2$-minute
noise pulse of strength $D=1.0$ is used to drive the concentration
to $\sim 2.2~\mu$M. Following this pulse, the noise is returned to
its original value. At $80$ minutes, a smaller $10$-minute noise
pulse of strength $D=0.1$ is used to return the concentration to
near its original value. The simulation technique is that of
Ref.~\cite{sancho}.
\\
\\
\noindent FIG.~2. The relaxation oscillator.  (A) Schematic of the
circuit.  The $P_{RM}$ promoter is used on two plasmids to control
the production of repressor~(X) and RcsA~(Y). After dimerization,
repressor acts to turn on both plasmids through its interaction at
$P_{RM}$.  As its promoter is activated, RcsA concentrations rise,
leading to an induced reduction of repressor. (B) Simulation of
Eqs.~(\ref{nodimxy}).  Oscillations arise as the RcsA-induced
degradation of repressor causes a transversal of the hysteresis
diagram in Fig.~1A.  The parameter values are $m_x=10$, $m_y=1$,
$\gamma_x=0.1$, $\gamma_y=0.01$, and $\gamma_{xy}=0.1$.
\\
\\
\noindent FIG.~3. Oscillatory regimes for the relaxation
oscillator. (A) The bifurcation wedge is larger for smaller values
of the parameter $\gamma_{xy}$. This larger regime corresponds to
larger values of the RcsA degradation parameter $\gamma_y$. Note
that the native~(i.e., without tuning) degradation rates of $
\gamma_x \sim \gamma_y \sim 0.005$ are very near the oscillatory
regime. (B) Bifurcation diagrams as a function of $\gamma_x$ and
$\gamma_{xy}$, and for two fixed values of $\gamma_y$. The
oscillatory regime is increased for smaller values of $\gamma_y$,
and, in both cases, small values of $\gamma_x$ are preferable for
oscillations. (C) The bifurcation diagram as a function of the
copy numbers $m_x$ and $m_y$, and for fixed degradation rates.
Importantly, one can adjust the periodic regime to account for the
unknown parameter $\gamma_{xy}$. The figure also indicates that,
for oscillations, one should choose as large a copy number as
possible for the plasmid containing the repressor protein~($m_x$).
In (A) and (B), constant parameter values are $m_x=10$ and
$m_y=1$, and in (C) $\gamma_x=1.0$ and $\gamma_y=0.01$.
\\
\\
\noindent FIG.~4. Parameter dependence of the oscillatory period.
(A) An increase in $\gamma_y$ decreases the period of
oscillations. (B) The period depends very weakly on the copy
number. In (A) $m_x=10$ and in (B) $\gamma_y=0.01$, and for both
plots, other parameter values are $\gamma_x=0.1$, $\gamma_xy=0.1$,
and $m_y=1$.
\\
\\
\noindent FIG.~5. Dynamics of a periodically driven relaxation
oscillator - Eq.(15); $\gamma_x=0.1$, $\gamma_y=0.01$,
$\gamma_{xy}=0.1$. (A) Resonant regions in period-amplitude
parametric plane. Solid lines (limit lines of periodic solutions)
together with dashed (period doubling) lines define boundaries of
stable periodic solutions for a given phase locking region M:N,
where M is the number of relaxation oscillation and N is the
number of driving sinusoidal oscillations. (B,C,D) Oscillations in
periodically driven repressor~(top curve) concentration together
with the oscillations of the sinusoidal driving~(bottom curve).
(B) 1:1 synchronization; the 14.6 minute period of cI oscillations
is equal to the driving period. (C) 1:2 phase locking; the 29.2
minute period of the cI oscillations is twice as long as the
driving period. (D) 2:1 phase locking; the 7.6 minute period of
the cI oscillations is equal to one half of the driving period.
\\
\\
\noindent FIG.~6. (A) As a function of the driving amplitude
$\Gamma$, the amplitude $\Gamma_x$ of the induced network
oscillations shows a sharp increase for a critical value of the
drive. The critical value corresponds to a drive large enough to
induce the hysteretic oscillations, and it increases as one
decreases $\gamma_y$ and moves away from the oscillatory region in
parameter space. The three curves denoted 1,2, and 3 are for
$\gamma_y$ values of 0.0038, 0.0036, and 0.0034. (B) The gain as a
function of the drive amplitude for $\gamma_y=0.0038$. Close to
the oscillatory region, a significant gain in the drive amplitude
can be induced. Parameter values for both plots are
$\gamma_x=0.1$, $\gamma_{xy}=0.1$, $m_x=10$, and $m_y=1$. Note
that, corresponding to these values, the network does not
oscillate (without driving) for $\gamma_y < 0.004$~(see the bottom
wedge of Fig.~3A).
\\
\\
\noindent FIG.~7 Null-clines for the two-protein bistable switch
systems. Stable fixed points are marked with circles, and unstable
fixed points are marked with squares. (A) Null-clines for the
synthetic \( \lambda  \) switch, Eqs.~(\ref{Eqn:x_dot}). Solid
line: \( \dot{x}=0 \) cline. Dashed line: \( \dot{y}=0 \) cline.
Parameter values: \( \gamma _{x}=0.004 \); \( \gamma _{y}=0.008
\); \( \rho _{y}=62.92 \); \( \alpha =11 \); \( m_{x}=m_{y}=1 \);
\( \sigma _{1}=2 \); \( \sigma _{2}=0.08 \); \( \beta _{1}=\beta
_{2}=\beta _{3}=0.08 \); and \( \beta _{4}=\beta _{5}=1 \). (B)
Null-clines for the toggle switch, Eqs.~(\ref{Eqn:u_dot}). Solid
line: \( \dot{u}=0 \) cline. Dashed line: \( \dot{v}=0 \) cline.
Parameter values (from Ref.~\cite{gardner}): \( \alpha _{1}=156.25
\); \( \alpha _{2}=15.6 \); \( \delta =2.5 \); \( \mu =1 \); \(
\eta =2.0015 \); \( [IPTG]=0 \); \( k=1 \).
\\
\\
\noindent FIG.~8. Transitions between stable states for the
two-protein bistable switch systems. The protein concentrations
have been normalized (the trace for each protein is normalized
relative to its own maximum value). The system parameters are
varied over time, altering the stability of the system and causing
transitions, as described in the text. Upper Plots: The switching
of Lac~(solid) and CI~(dashed) in the synthetic toggle
model~\cite{gardner}. The parameter values are as given in the
caption to Fig.~7, except as follows. (1--4 hours): $[IPTG]=2$ mM,
$k=1.0$. (7--10 hours): $[IPTG]=0.0$, $k=50.81$. Lower Plots: The
switching of CRO~(dashed line) and CI~(solid) in the synthetic
$\lambda$ model. The parameter values are as given in the caption
to Fig.~7, except as follows. (1--4 hours): $\gamma _{x}=0.004$,
$\gamma _{y}=21.6$. (7--10 hours): $\gamma_{x}=18.0$,
$\gamma_{y}=0.008$.


 \newpage
 \clearpage
 \pagestyle{empty}
 ~\\~\\~\\~\\
 \begin{center}
 \epsfig{file=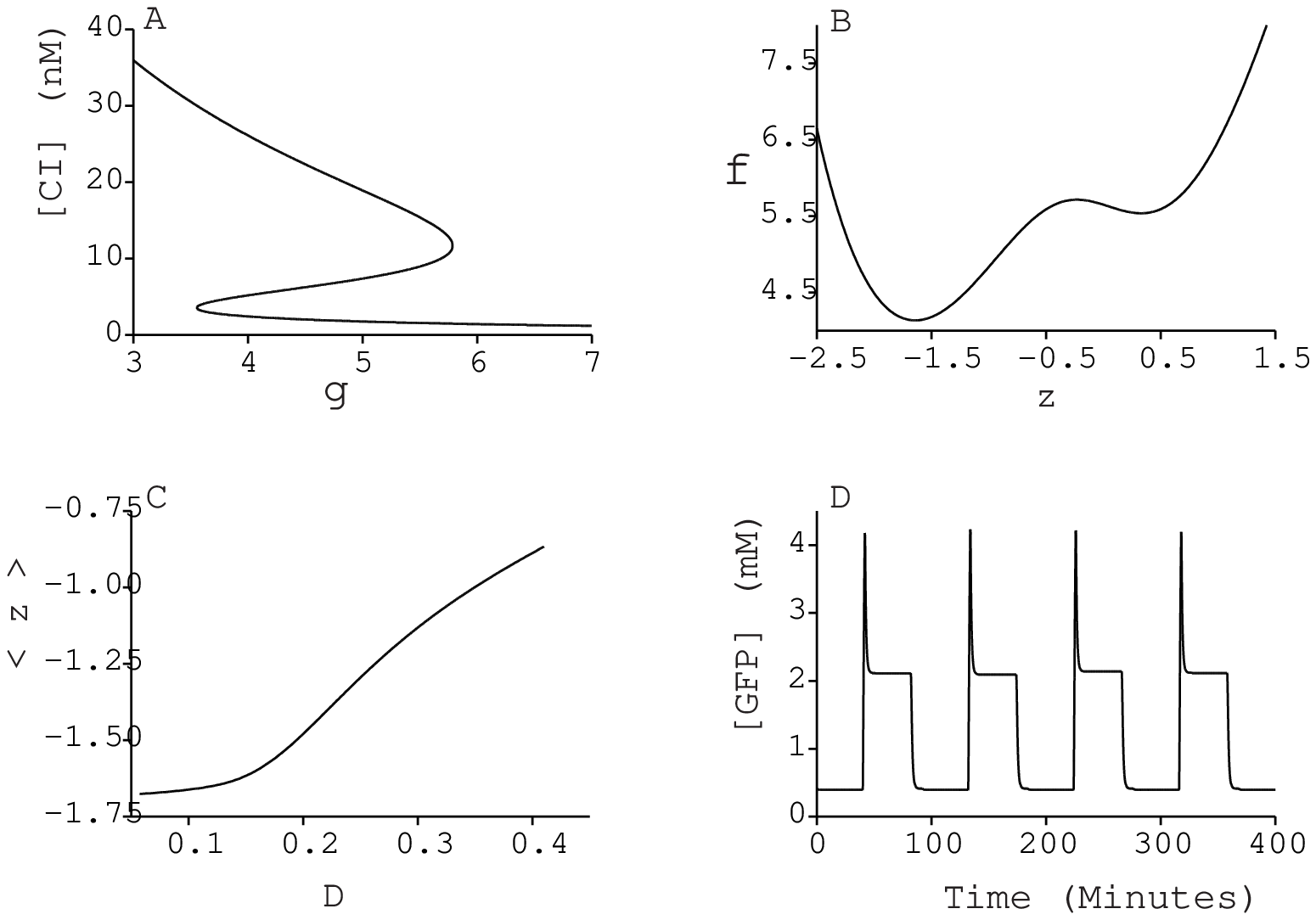,height=4in}
~\\~\\~\\~\\~\\
  Figure 1 - Hasty et al.
 \end{center}

 \newpage
 \clearpage
 \pagestyle{empty}
  ~\\~\\~\\~\\~\\
 \begin{center}
  \epsfig{file=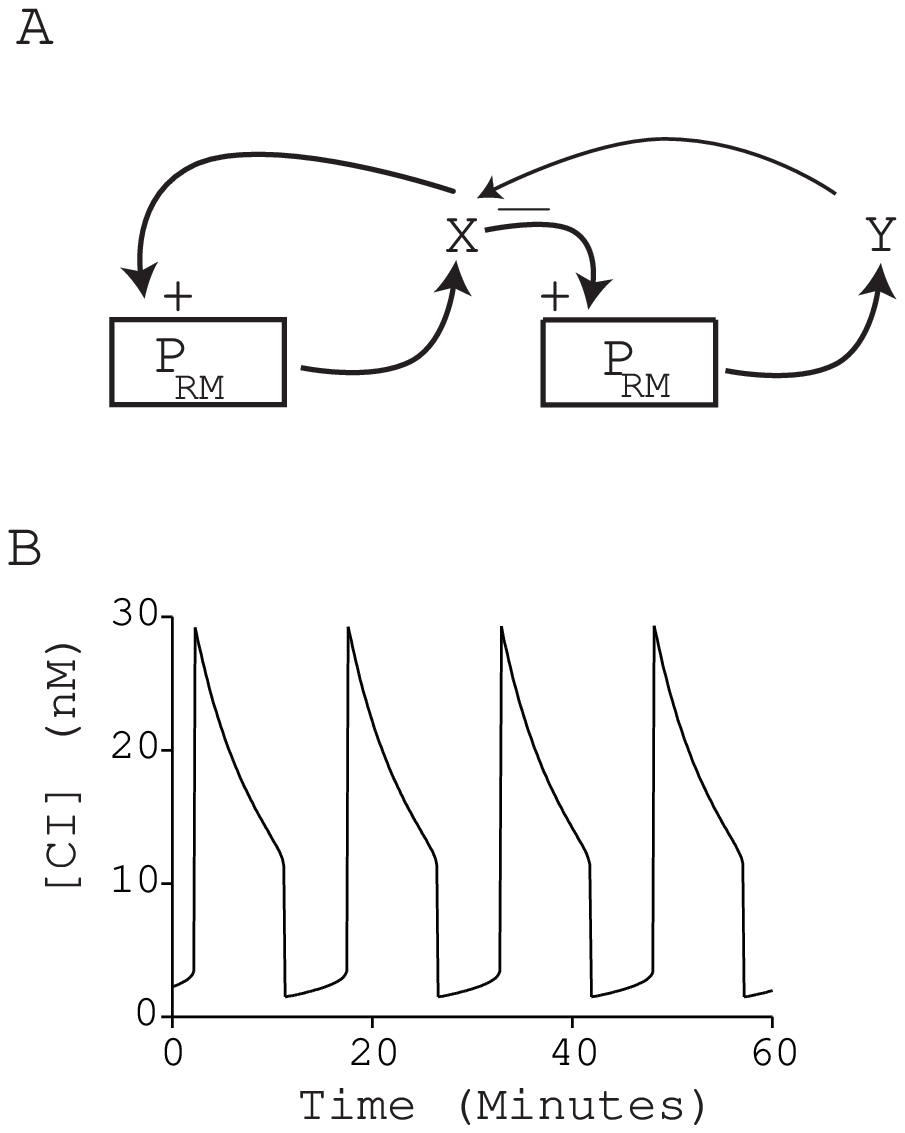,height=4in}
  ~\\~\\~\\~\\~\\~\\~\\~\\~\\
 Figure 2 - Hasty et al.
 \end{center}

 \newpage
 \clearpage
 \pagestyle{empty}
  ~\\~\\~\\
 \begin{center}
 \epsfig{file=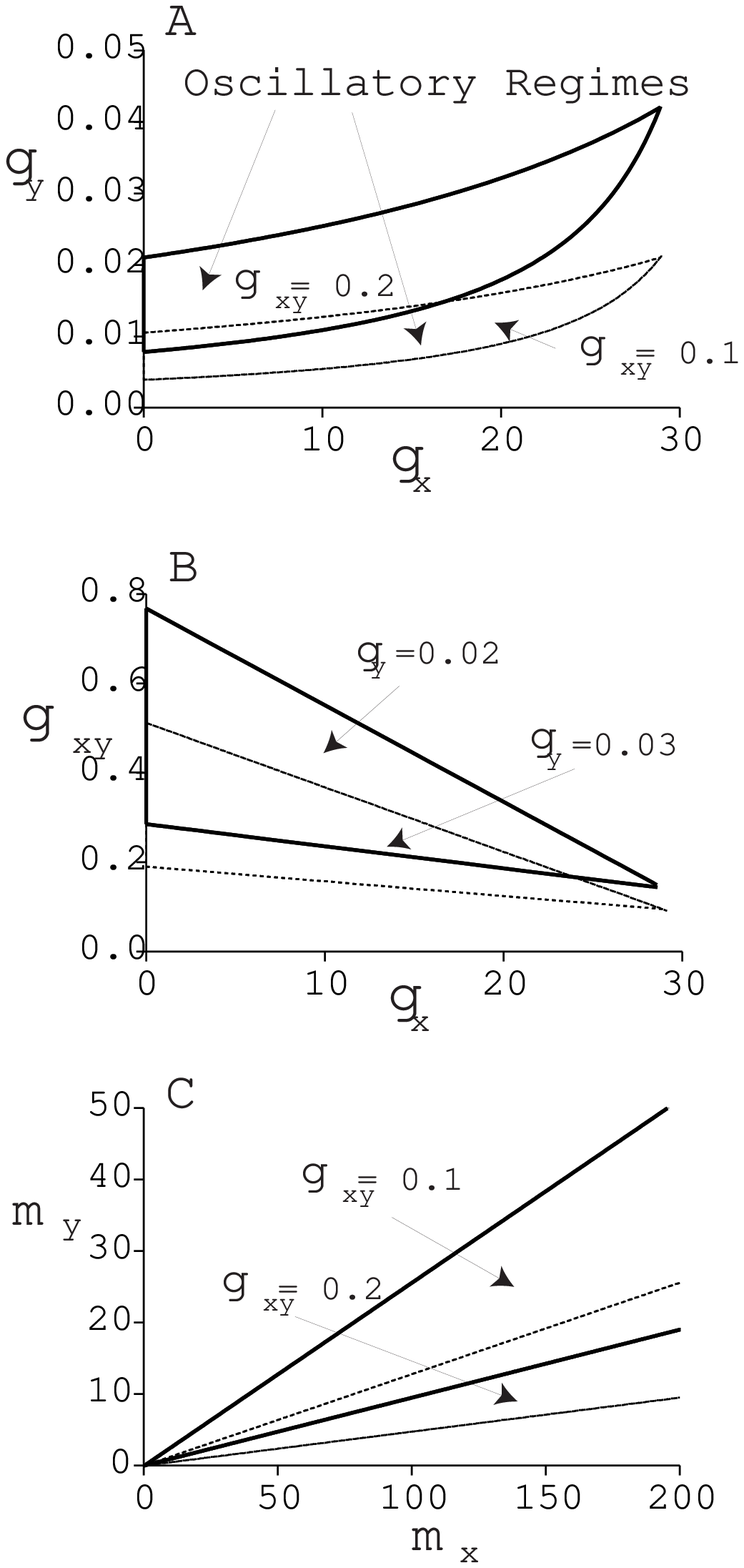,width=3in}
 ~\\~\\~\\
 Figure 3 - Hasty et al.
 \end{center}

 \newpage
 \clearpage
 \pagestyle{empty}
  ~\\~\\~\\
 \begin{center}
 \epsfig{file=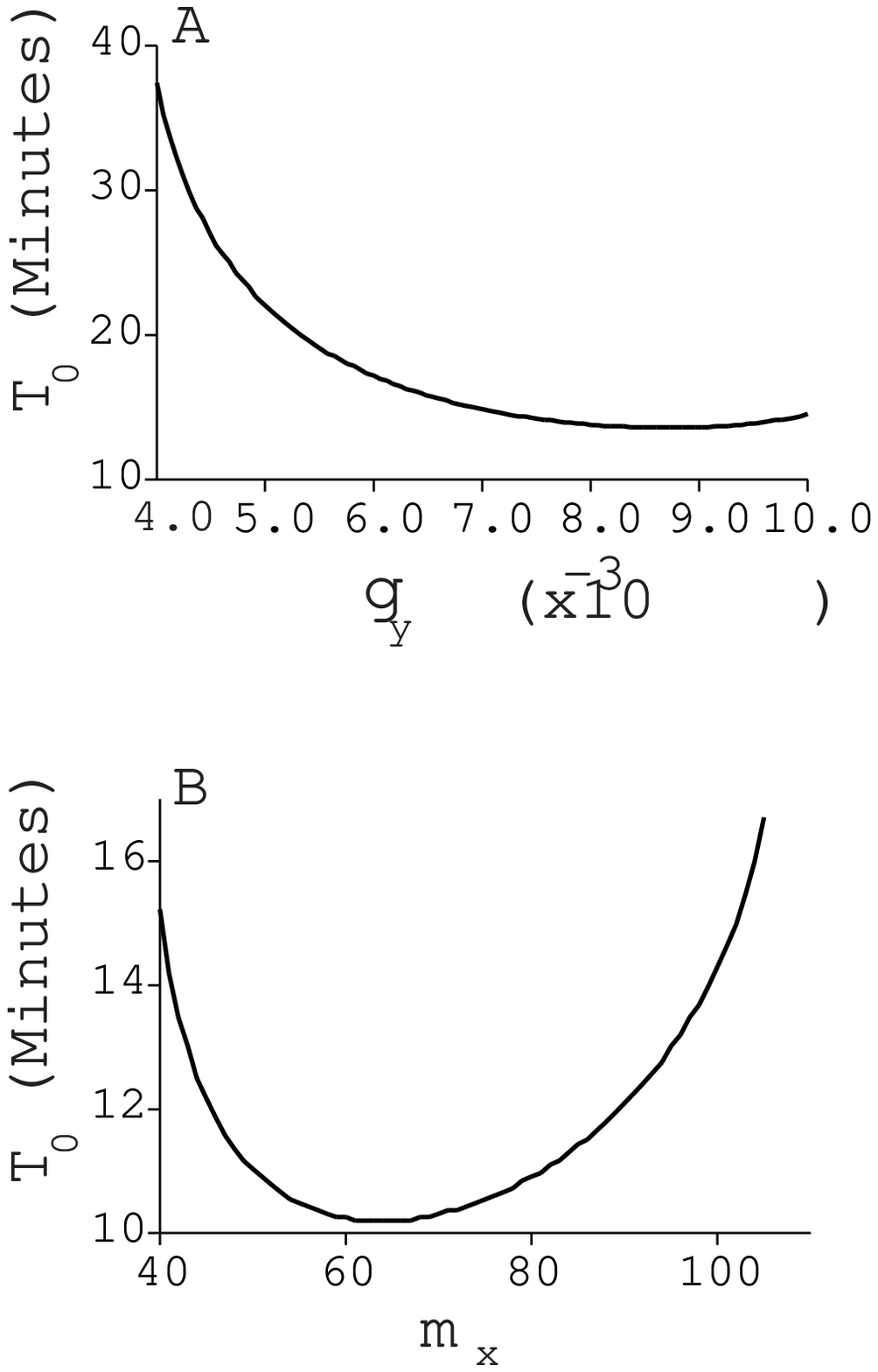,width=3in}
  ~\\~\\~\\~\\~\\~\\~\\~\\~\\
 Figure 4 - Hasty et al.
 \end{center}

  \newpage
 \clearpage
 \pagestyle{empty}
 ~\\~\\~\\~\\
 \begin{center}
 \epsfig{file=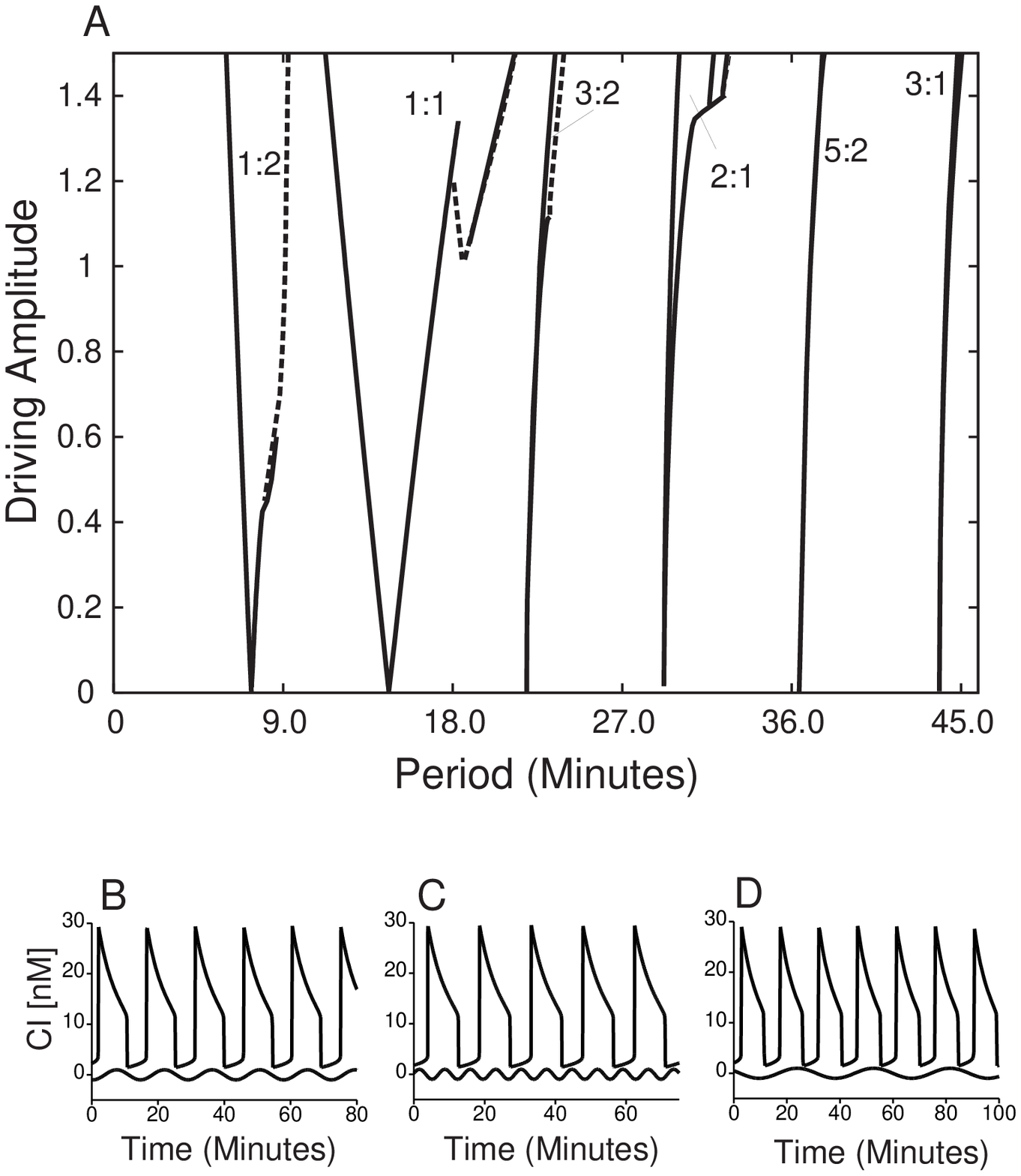,height=4in}
~\\~\\~\\~\\~\\~\\~\\
  Figure 5 - Hasty et al.
 \end{center}

 \newpage
 \clearpage
 \pagestyle{empty}
  ~\\~\\~\\
 \begin{center}
 \epsfig{file=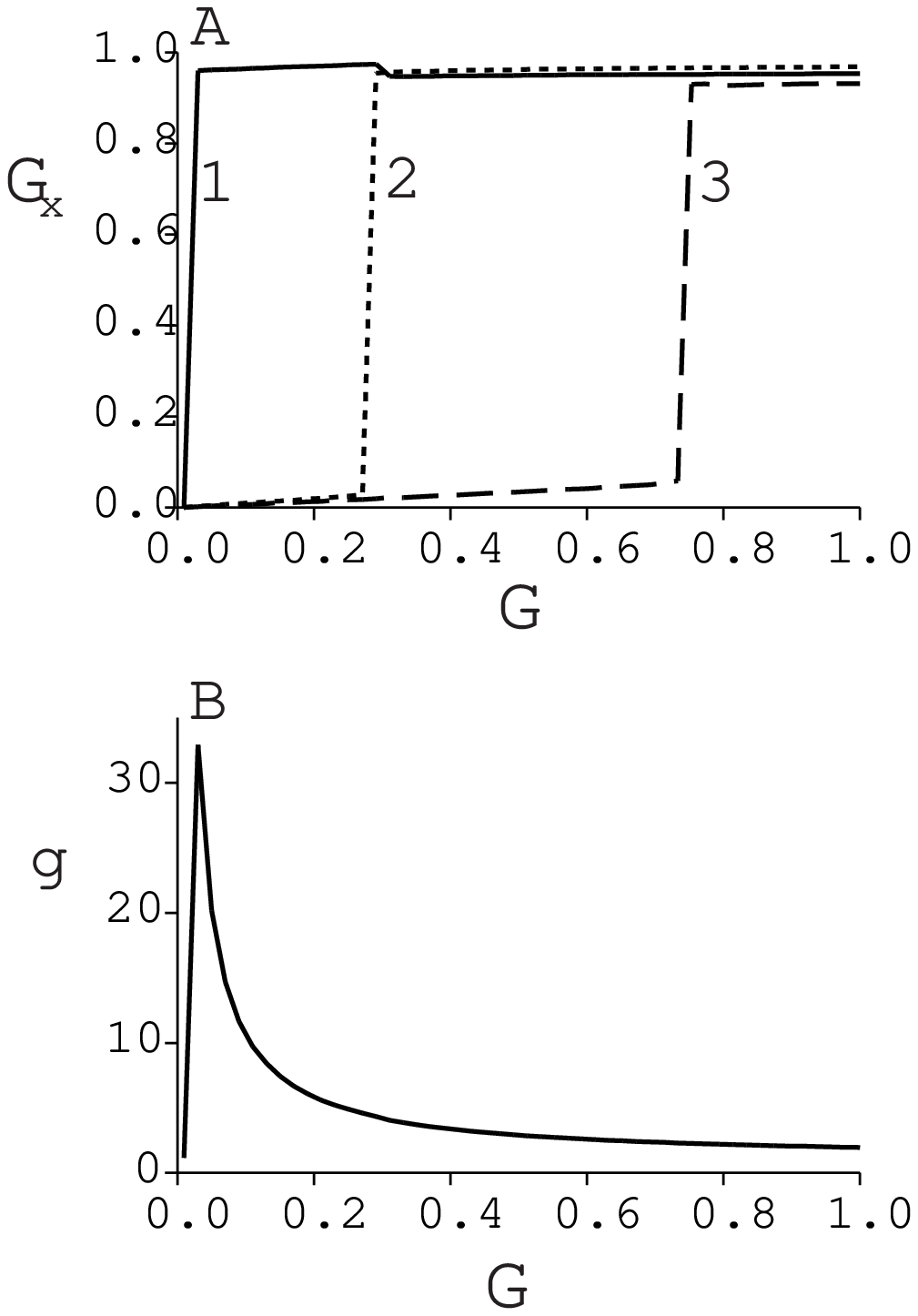,width=3in}
  ~\\~\\~\\~\\~\\~\\~\\~\\~\\
 Figure 6 - Hasty et al.
 \end{center}

 \newpage
 \clearpage
 \pagestyle{empty}
 ~\\~\\~\\~\\
 \begin{center}
 \epsfig{file=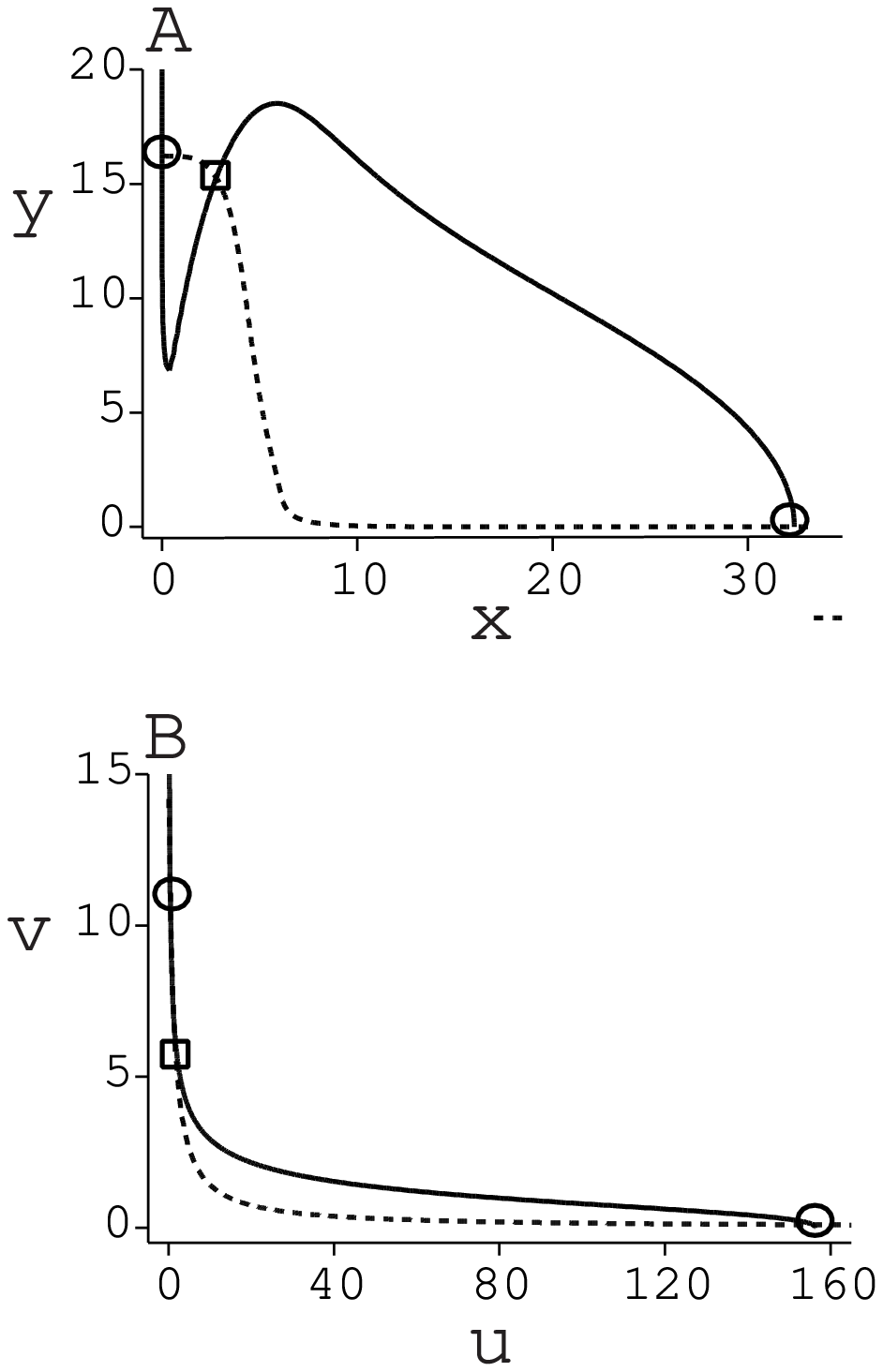,height=4in}
~\\~\\~\\~\\~\\~\\~\\
  Figure 7 - Hasty et al.
 \end{center}

  \newpage
 \clearpage
 \pagestyle{empty}
 ~\\~\\~\\~\\
 \begin{center}
 \epsfig{file=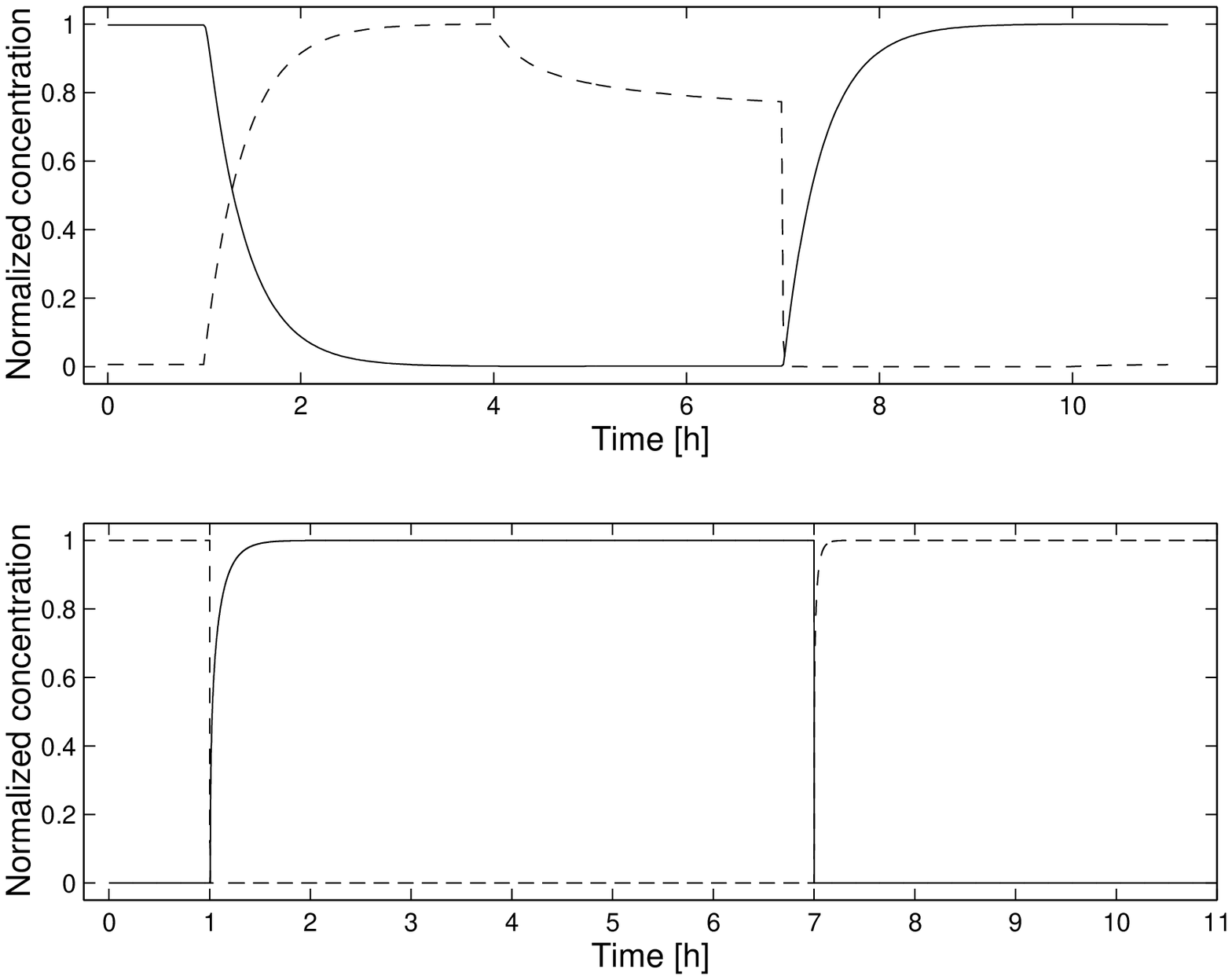,height=4in}
~\\~\\~\\~\\~\\~\\~\\
  Figure 8 - Hasty et al.
 \end{center}


\begin{thebibliography}{1}
\bibitem{footnote_lambda} A more general treatment would
incorporate all possible DNA-repressor binding configurations.
However, the additional reactions have binding affinities
characterized by equilibrium constants that are significantly
smaller than for those considered, and their inclusion will not
alter the general conclusions derived in this work.
\end{thebibliography}
\end{document}